\documentclass{PoS}

\usepackage{graphicx}
\usepackage{amsmath}
\usepackage{xcolor}
\newcommand{\Tr}{Tr}

\newcommand{\dt}{dt}

\title{Spectroscopy With Local Multi-hadron Interpolators In Lattice QCD}

\ShortTitle{Spectroscopy With Multi-hadron Interpolators In Lattice QCD}

\author{\speaker{Adrian L. Kiratidis}${}^{a}$, Waseem Kamleh${}^a$, Derek B. Leinweber${}^a$, Zhan-Wei Liu${}^{a,b}$,  Finn M. Stokes${}^a$, Anthony W. Thomas${}^{a,c}$\\
${}^a$Special Research Center for the Subatomic Structure of Matter (CSSM), Department of Physics,
University of Adelaide, Adelaide, South Australia 5005, Australia
\\
${}^b$School of Physical Science and Technology, Lanzhou University, Lanzhou 730000, China\\
${}^c$ARC Centre of Excellence in Particle Physics at the Terascale, Department of Physics, University of Adelaide,
Adelaide, South Australia 5005, Australia\\
E-mail: \email{adrian.kiratidis@adelaide.edu.au}}

\abstract{The positive-parity nucleon spectrum is studied in 2 + 1 flavour lattice QCD in an attempt to discover novel low-lying energy eigenstates in the region of the Roper resonance.  In this work, we employ standard three-quark interpolating fields and introduce new local five-quark meson-baryon operators that hold the possibility of revealing new states that have been missed in previous studies.  Motivated by phenomenological arguments, five-quark interpolators based on the $\sigma{N}$, $\pi{N}$ and $a_0{N}$ channels are constructed. Despite the introduction of qualitatively different operators, no novel energy levels are extracted near the regime of the Roper resonance.}

\FullConference{The 26th International Nuclear Physics Conference\\
                 11-16 September, 2016\\
                 Adelaide, Australia}

\begin{document}

\section{Introduction}
\label{sect:Introduction}
In recent times, considerable resources have been invested into studying the nucleon spectrum on the lattice, both to elucidate the properties of
various excitations, and to probe the validity of the calculation methodology itself.  In particular, the first positive-parity $J^P = {\frac{1}{2}}^{+}$ excitation of the nucleon, known as the
Roper resonance $N^{*}(1440)$, has been a long-standing puzzle for the community.  In constituent quark models, the Roper resonance energy level resides above the lowest-lying negative-parity excitation, the $N^{*}(1535)$, while in Nature, the Roper lies $95\textrm{ MeV}$ below the $N^{*}(1535)$.

A crucial component for lattice spectroscopy studies in this channel is to select a sufficiently large operator basis so as to appropriately span the states of interest in the low-lying spectrum.  One method for performing this selection is to include qualitatively different
operators, by constructing interpolators with different quark and/or Dirac structure while retaining the quantum numbers relevant to the channel.  

As the spectrum has many multi-particle scattering states present, including five-quark interpolating fields is one natural way to construct these novel operators.  Their inclusion then enables us to search for states that have poor overlap with standard three-quark interpolators and have been missed.  As meson-baryon states with strong attraction can be well localized~\cite{Liu:2016uzk}, local five-quark operators can be expected to overlap well with such localized states.
The presence of such states would lead to an important role for molecular meson-baryon 
configurations~\cite{Hall:2014uca} in our understanding of the Roper resonance.

In this work we consider pairing positive-parity meson operators with the standard nucleon interpolating field in order to allow the ground state nucleon to participate in forming the quantum numbers of the Roper.
To this end, we construct local five-quark meson-baryon operators $\sigma{N}$ and $a_{0}{N}$,
and explore their impact on the positive-parity nucleon spectrum.  We are searching for any alteration of the spectrum reported in previous studies and/or any new low-lying finite volume energy eigenstates that have hitherto remained elusive.

Following the outline of variational analysis techniques in
Section~\ref{sect:CorrelationMatrixTechniques}, we construct the new local five-quark operators in
Section~\ref{sect:InterpolatingFields}.  Simulation details are discussed in
Section~\ref{sect:Simulation Details} and the results of the variational analyses are presented in
Section~\ref{sect:Results}.  A summary of our findings and their impact our our understanding of
the Roper resonance is presented in Sec.~\ref{sect:Conclusions}.
%    
%%%%%%%%%%%%%%%%%%%%%%%%%%%%%%%%%%%%%%%%%%%%%%
%
\section{Correlation Matrix Techniques}
\label{sect:CorrelationMatrixTechniques}

Correlation matrix based variational analyses~\cite{Michael:1985ne, Luscher:1990ck} are well-established as a methodology within which hadron spectra can be extracted~\cite{Leinweber:2015kyz}. The calculation begins via the judicious selection of a suitably large
basis of $N$ operators, such that the states of interest within the spectrum are contained within
the span.  An $N \times N$ matrix, $G_{ij}$, of cross correlation functions,
\begin{equation}
\label{defn:CM}
{G}_{ij}(\vec{p},t) = \Tr\left( \Gamma_{\pm}\sum_{\vec{x}}\textrm{e}^{-i\vec{p}\cdot\vec{x}}\,\big\langle\,\Omega\, \big| \,\chi_{i}(\vec{x},t)\,\overline{\chi}_{j}(\vec{0},t_{src})\, \big| \,\Omega\, \big\rangle\right) = \sum_{\alpha}\lambda^{\alpha}_{i}\,\bar{\lambda}^{\alpha}_{j}\,\textrm{e}^{-m_{\alpha}t},
\end{equation}
where the operator $\Gamma_{\pm} = \frac{1}{2}\,(\gamma_{0} \pm I)$ projects out a definite parity at $\vec{p}=\vec{0},$ $\lambda^{\alpha}_{i}$ and $\bar{\lambda}^{\alpha}_{j}$ are the couplings of the annihilation, 
$\chi_{i}$, and creation, $\overline{\chi}_{j}$, operators at the sink and source respectively, and $m_{\alpha}$ enumerates (in $\alpha$) the energy eigenstates of mass $m_{\alpha}$.  We then search for an appropriate
linear combination of operators
\begin{equation}
\bar{\phi}^{\alpha} = \bar{\chi}_{j}\,u^{\alpha}_{j} \qquad \textrm{ and } \qquad \phi^{\alpha} = \chi_{i}\,v^{\alpha}_{i}
\end{equation}  
such that $\phi$ and $\bar{\phi}$ couple solely to a single energy eigenstate.  It can then be shown that for a given choice of variational parameters $(t_0,\dt)$, $u^{\alpha}_{j}$ and
$v^{\alpha}_{i}$ can be obtained by solving the left and right eigenvalue equations
\begin{align}
\label{E-value-eq-L}
\big[{G}^{-1}(t_{0})\,{G}(t_{0} + \dt)\big]_{ij}\,u^{\alpha}_{j} &= c^{\alpha}\,u^{\alpha}_{i}\\
\label{E-value-eq-R}
v^{\alpha}_{i}\,\big[{G}(t_{0} + \dt)\,{G}^{-1}(t_{0})\big]_{ij} &=
c^{\alpha}\,v^{\alpha}_{j},
\end{align}
with eigenvalue $c^{\alpha} = \textrm{e}^{-m_{\alpha}\dt}$.

These eigenvectors $u^{\alpha}_{j}$ and
$v^{\alpha}_{i}$ diagonalise the correlation matrix at $t_{0}$ and $t_{0} +\dt$, which enables us to write down the
eigenstate-projected correlation function as

\begin{equation}
{G}^{\alpha}(t) = v^{\alpha}_{i}\,{G}_{ij}(t)\,u^{\alpha}_{j}.
\end{equation}
which is used to extract masses.  Further technical details that differentiate our method from other methods, along with an outline of our fitting technique can be found in Refs.~\cite{Kiratidis:2015vpa} and~\cite{Kiratidis:2016hda}.
%
%%%%%%%%%%%%%%%%%%%%%%%%%%%%%%%%%%%%%%%%%%%%%%%
%
\section{Interpolating Fields}
\label{sect:InterpolatingFields}

Previous
work with five-quark interpolators~\cite{Kiratidis:2015vpa, Lang:2012db}
have successfully extracted states consistent with multi-particle scattering
thresholds in the negative-parity nucleon channel.  Motivated by this success, we employ a similar tactic constructing five-quark operators in the positive-parity channel.

Making use of the operators for the positive-parity isocsalar $\sigma$ and isovector $a^{0}_{0}$ and
$a^{+}_{0}$ mesons
\begin{align}
\sigma &= \frac{1}{\sqrt{2}}\big[\bar{u}^{e}\,I\,u^{e} + \bar{d}^{e}\,I\,d^{e}\big]\, , \nonumber\\
a^{0}_{0} &= \frac{1}{\sqrt{2}}\big[\bar{u}^{e}\,I\,u^{e} - \bar{d}^{e}\,I\,d^{e}\big]\, , \nonumber\\
a^{+}_{0} &= \big[\bar{d}^{e}\,I\, u^{e}\big]\, ,
\end{align}
we are able to construct five-quark $\sigma{N}$- and $a_{0}{N}$-type interpolators.
Naturally, we employ the labels $(\pi{N},\sigma{N},a_0{N})$ only to distinguish the mathematical structure of the states, as each of these operators couple to multiple states on the lattice.
The general form of the $\sigma{N}$-type operators is given by
\begin{align}
\chi_{\sigma{N}}(x) &= \frac{1}{2}\,\epsilon^{abc}\,\big[u^{Ta}(x)\,\Gamma_{1}\,d^{b}(x)\big]\,\Gamma_{2}\,u^{c}(x)\nonumber\\
&\qquad\qquad\times\big[\bar{u}^{e}(x)\,I\,u^{e}(x) + \bar{d}^{e}(x)\,I\,d^{e}(x)\big].
\end{align}
The choices of $(\Gamma_{1}, \Gamma_{2}) = (C\gamma_{5}, \textrm{I})$ and $(C,\gamma_{5})$ then
provide us with two five-quark operators $\chi_{\sigma{N}}(x)$ and $\chi^{\prime}_{\sigma{N}}(x)$
respectively.

Similarly, using the Clebsch-Gordan coefficients to project isospin $I = 1/2$, $I_{3} = +1/2$ the general form of the $a_{0}{N}$-type interpolators is given by
\begin{align}\label{Proton5QrkOpInterpolator}
\chi_{a_0{N}}(x) &= \frac{1}{\sqrt{6}}\,\epsilon^{abc}\times\nonumber\\
&\quad\bigg\{2\big[u^{Ta}(x)\,\Gamma_{1}\,d^{b}(x)\big]\,\Gamma_{2}\,d^{c}(x)\,\big[\bar{d}^{e}(x)\,I\,u^{e}(x)\big]\nonumber\\
&\quad - \big[u^{Ta}(x)\,\Gamma_{1}\,d^{b}(x)\big]\,\Gamma_{2}\,u^{c}(x)\,\big[\bar{d}^{e}(x)\,I\,d^{e}(x)\,\big]\nonumber\\
&\quad + \big[u^{Ta}(x)\,\Gamma_{1}\,d^{b}(x)\big]\,\Gamma_{2}\,u^{c}(x)\,\big[\,\bar{u}(x)^{e}\,I\,u^{e}(x)\big]\bigg\}.
\end{align}
Once again the two aforementioned choices of $(\Gamma_{1}, \Gamma_{2})$ provide $\chi_{a_0{N}}(x)$ and
$\chi^{\prime}_{a_0{N}}(x)$ respectively.  We also utilize the two five-quark operators $\chi_{\pi{N}}$ and $\chi^{\prime}_{\pi{N}}$ based on the form
\begin{align}\label{Proton5QrkOpInterpolator}
\chi_{\pi{N}}(x) &= \frac{1}{\sqrt{6}}\,\epsilon^{abc}\,\gamma_{5}\times\nonumber\\
&\quad\bigg\{2\big[u^{Ta}(x)\,\Gamma_{1}\,d^{b}(x)\big]\,\Gamma_{2}\,d^{c}(x)\,\big[\bar{d}^{e}(x)\,\gamma_{5}\,u^{e}(x)\big]\nonumber\\
&\quad - \big[u^{Ta}(x)\,\Gamma_{1}\,d^{b}(x)\big]\,\Gamma_{2}\,u^{c}(x)\,\big[\bar{d}^{e}(x)\,\gamma_{5}\,d^{e}(x)\,\big]\nonumber\\
&\quad + \big[u^{Ta}(x)\,\Gamma_{1}\,d^{b}(x)\big]\,\Gamma_{2}\,u^{c}(x)\,\big[\,\bar{u}(x)^{e}\,\gamma_{5}\,u^{e}(x)\big]\bigg\},
\end{align}
and detailed in Ref.~\cite{Kiratidis:2015vpa}. Our basis of interpolating fields is completed with the inclusion of the standard three-quark nucleon operators $\chi_{1}$ and
$\chi_{2}$ given by
\begin{align}\label{NucleonInterps}
\chi_{1} &= \epsilon^{abc}[u^{aT}\, (C\gamma_{5})\, d^{b}]\, u^{c}\nonumber\\
\chi_{2} &= \epsilon^{abc}[u^{aT}\, (C)\, d^{b}]\, \gamma_{5}\, u^{c}.
\end{align}

The introduction of five-quark operators necessarily introduces diagrams with loop propagator contributions.
Loop propagators at the source, $S(0,0)$, are a subset of standard point-to-all propagators via $S(x,0)|_{x=0}$ while loops at
the sink, $S(x,x)$, are estimated stochastically by averaging over four independent
$\mathbb{Z}_4$ noise vectors.  We employ full dilution in time, spin and colour.  Further details can be found in our previous work~\cite{Kiratidis:2015vpa}.

\section{Simulation Details}
\label{sect:Simulation Details}

Throughout this work we use the PACS-CS $2 + 1$ flavour dynamical-fermion configurations~\cite{Aoki:2008sm} that are made available through the the ILDG~\cite{Beckett:2009cb}.  There are $32^3 \times 64$ lattice points with a spacing of $0.0907\textrm{ fm}$ which provides a physical volume of $\approx (2.90\textrm{ fm})^3$.  The light quark mass is set by $\kappa_{ud} = 0.13754$ which leads to a pion mass of  $m_{\pi} = 411 \textrm{ MeV}$.  Gauge invariant Gaussian smearing~\cite{Gusken:1989qx} is used at the source and sink to alter the overlap of our operators with states with in the spectrum.  We employ $n_s = 35, 100$ and 200 sweeps of Gaussian smearing.  Further information about the gauge configurations and other relevant simulation details can be found in Ref.~\cite{Kiratidis:2016hda}.   

%%%%%%%%%%%%%%%%%%%%%%%%%%%%%%%%%%%%%%%%%%%%%%%%%%%%%%%%%%%%%
%
\section{Results}
\label{sect:Results}

\subsection{Correlation Matrix Construction}

\begin{figure}[!htbp]
  \centering 
  {\includegraphics[width=0.48\textwidth]{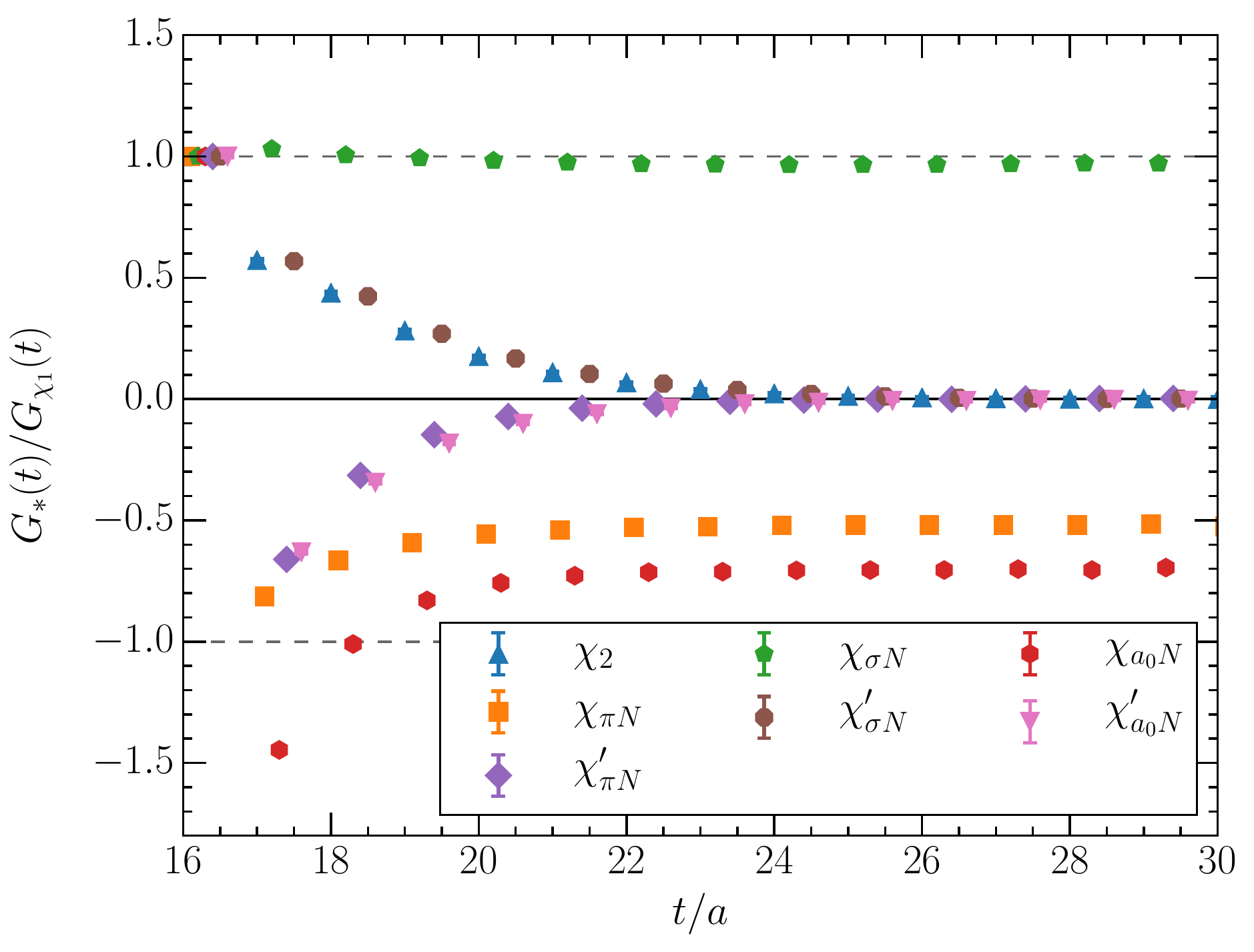}}
  {\includegraphics[width=0.48\textwidth]{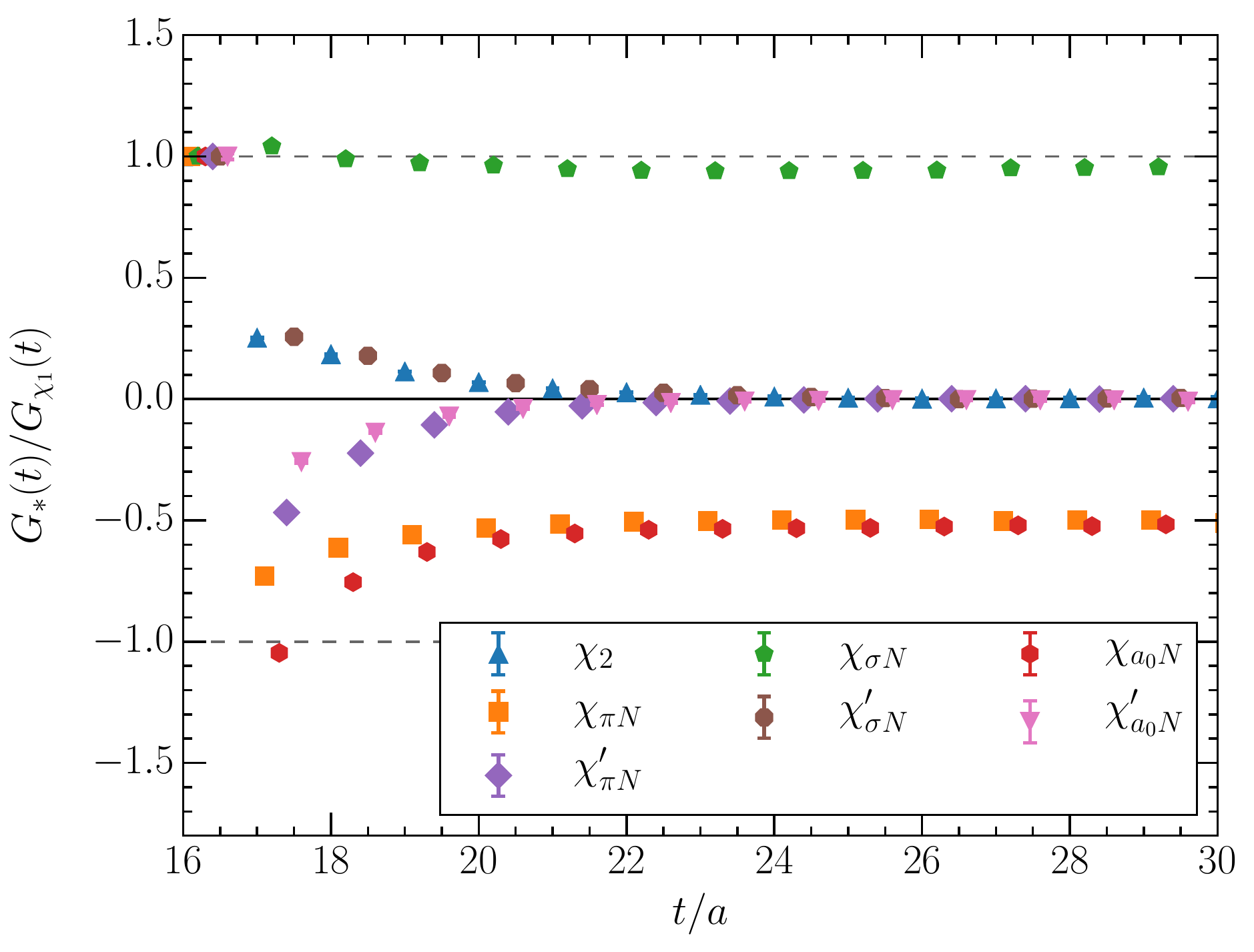}}\\
  {\includegraphics[width=0.48\textwidth]{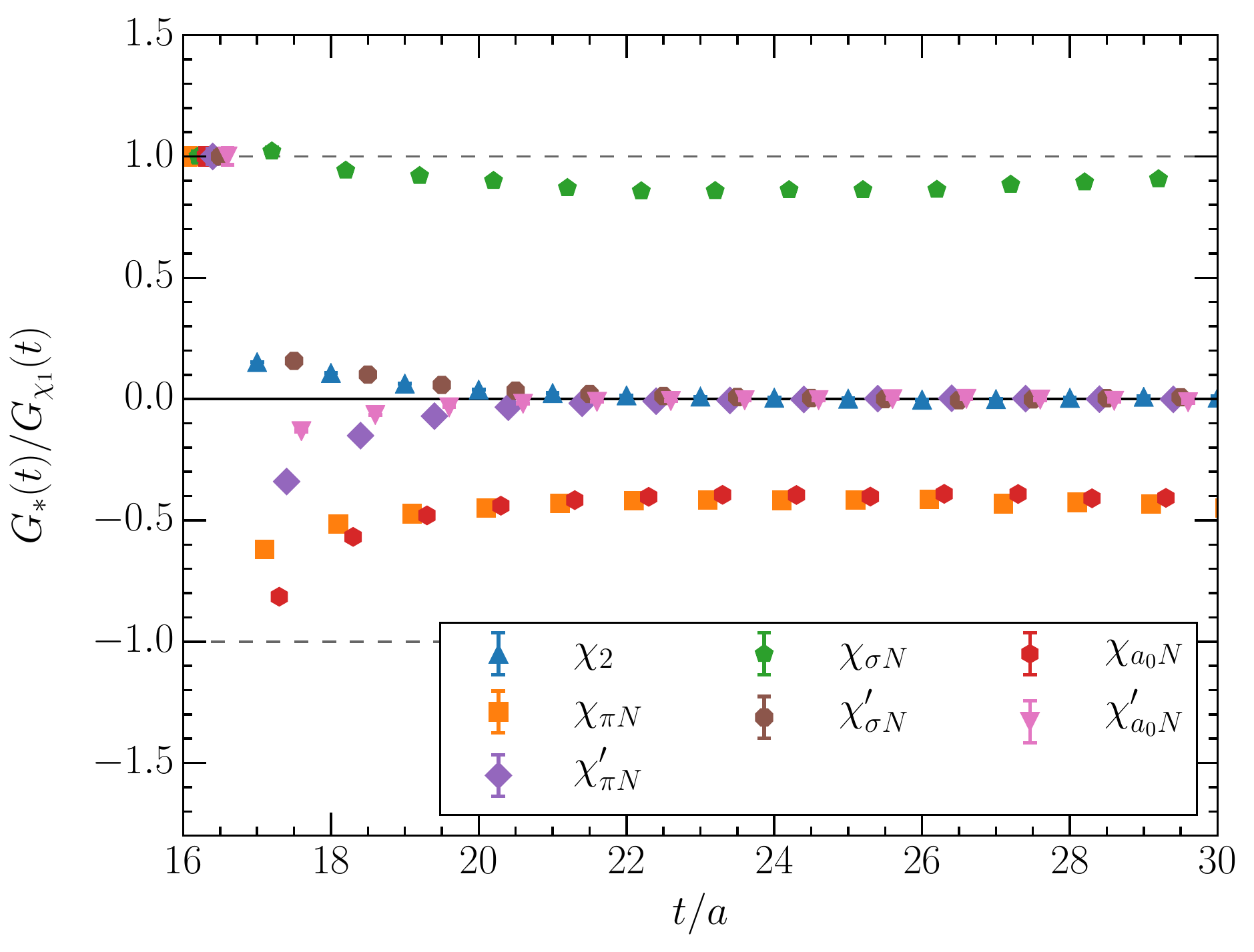}}
\caption{Correlation function ratios which are formed by dividing the correlator corresponding to each operator indicated in the legend by the correlation function 
  formed from the $\chi_{1}$ operator.  Plots are presented at 35 (left), 100 (right) and 200
  (bottom) sweeps of Gaussian smearing in the quark-propagator source and sink.  The $t$ component of the ratio has been sequentially offset for clarity.}
\label{fig:CorrelatorRatioComparison}
\end{figure}

In Section \ref{sect:InterpolatingFields} we constructed eight qualitatively distinct interpolators, and we employ three different levels of Gaussian smearing on them as discussed in Section \ref{sect:Simulation Details}.  Consequently, we posses twenty-four operators with which to build various correlation matrices.  In order to investigate which possible sub-bases may be instructive to study, we present a plot of correlation function ratios in Figure \ref{fig:CorrelatorRatioComparison}.  This will aid in identifying correlators that display a unique plateau approach, indicating overlap with a novel superposition of states.

Interestingly, the ratios formed from the $\sigma{N}$ type operators, that is
${G}_{\chi_{\sigma{N}}}/{G}_{\chi_{1}}$ and ${G}_{\chi^{\prime}_{\sigma{N}}}/{G}_{\chi_{1}}$ display remarkably similar behaviour to the ratios ${G}_{\chi_{1}}/{G}_{\chi_{1}}$ and
${G}_{\chi_{2}}/{G}_{\chi_{1}}$.
We therefore anticipate the overlap of states in the spectrum with the $\chi_{\sigma{N}}$ operator to be much the same as with the $\chi_{1}$ operator and similarly $\chi^{\prime}_{\sigma{N}}$ to posses a similar overlap with excitations as $\chi_{2}$.  Evidently, the novel $\sigma{N}$-type operators provide little new
information from that already present in $\chi_{1}$ and $\chi_{2}$, which comes as no surprise given the $\sigma$
meson has the quantum numbers of the vacuum.  Consequently, we omit the $\chi_{\sigma{N}}$ and $\chi^{\prime}_{\sigma{N}}$ operators from bases in which the matching $\chi_{1}$ or $\chi_{2}$ operator is also present.

Of the new $a_{0}{N}$-type operators, the ratio formed with $\chi_{a_{0}{N}}$ displays the most different approach to the plateau, and we therefore expect it to be the most promising operator to reveal an alteration in the low-lying spectrum when compared to previous analyses.

\begin{table}[t]
\caption{The operators used in constructing each correlation-matrix basis.}
\begin{center}
%\begin{ruledtabular}
\begin{tabular}{cl}
    Basis Number & Operators Used  \\[2pt]
    \hline 
\noalign{\vspace{2pt}}
    1 & $\chi_{1}$, $\chi_{2}$\\
    2 & $\chi_{1}$, $\chi_{2}$, $\chi_{a_0{N}}$\\
    3 & $\chi_{1}$, $\chi_{2}$, $\chi_{a_0{N}}$, $\chi^{\prime}_{a_0{N}}$\\
    4 & $\chi_{\pi{N}}$, $\chi^{\prime}_{\pi{N}}$, $\chi_{a_0{N}}$\\
    5 & $\chi_{\pi{N}}$, $\chi^{\prime}_{\pi{N}}$, $\chi_{a_0{N}}$, $\chi^{\prime}_{a_0{N}}$\\
    6 & $\chi_{\pi{N}}$, $\chi^{\prime}_{\pi{N}}$, $\chi_{\sigma{N}}$, $\chi^{\prime}_{\sigma{N}}$\\
    7 & $\chi_{\sigma{N}}$, $\chi^{\prime}_{\sigma{N}}$, $\chi_{a_0{N}}$, $\chi^{\prime}_{a_0{N}}$\qquad\null\\[2pt]
\end{tabular}
%\end{ruledtabular}
\end{center}
\label{table:BasisTable}
\end{table}

In order to choose a basis of operators sufficiently small so as to readily obtain a solution, we focus on sub-bases formed from correlators possessing 35 and 100 sweeps of smearing in the propagator sources and sinks, as these are the smearing levels that provide the most variation at
early times.  Although we will not detail the results of bases with 200 sweeps of smearing, the energy levels extracted were consistent with those presented herein.  We investigate seven distinct sub-bases, all formed with 35 and 100 sweeps or smearing and outlined in Table \ref{table:BasisTable}.

\subsection{Finite Volume Spectrum of States}

\begin{figure}[t]
  \centering 
  {\includegraphics[width=0.48\textwidth]{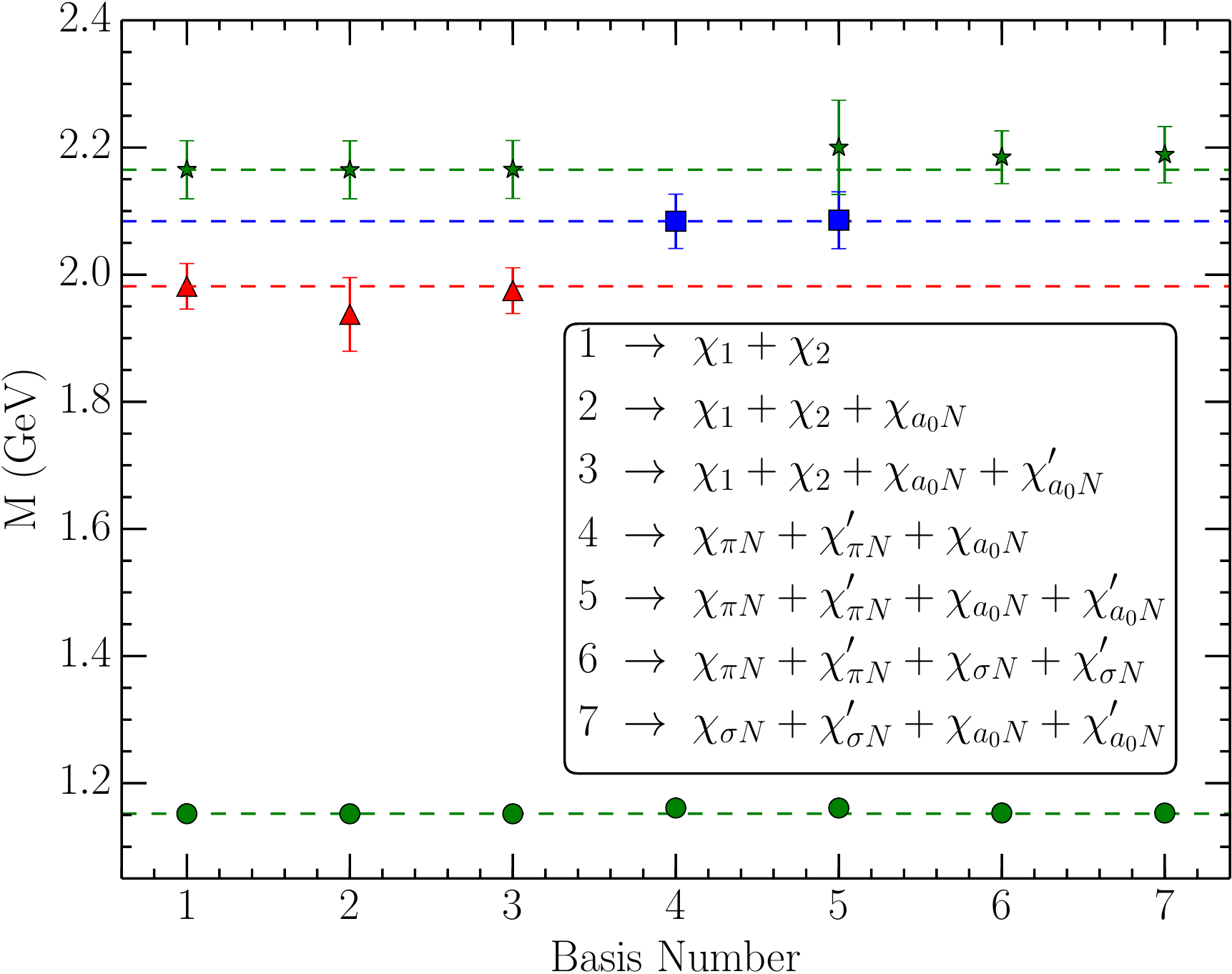}}\\
\caption{The low-lying spectra obtained from each of the correlation-matrix bases studied. For each operator, two smearing levels of $n_{s} = 35$ and $n_{s} = 100$ in all bases.  Dashed horizontal lines are present to guide the eye, and have been set by the central values from basis 1 in all cases except for the state $\sim 2.1$ GeV, in which case it is drawn from basis 4.}
\label{fig:SpectraPlot}
\end{figure}

In Fig.~\ref{fig:SpectraPlot} we present the the low-lying spectra obtained from the correlation matrices built from the bases detailed in Table~\ref{table:BasisTable}.
In basis number one, we display the results from a 4 $\times$ 4 analysis with the three-quark
$\chi_{1}$ and $\chi_{2}$ interpolating fields at two different smearing levels.  This basis can then be used as a benchmark to determine whether the introduction of five-quark operators alters the low-lying spectrum. 

We proceed via the introduction of the operators $\chi_{a_{0}N}$ and $\chi^{\prime}_{a_{0}N}$ in bases two and three respectively.  This reveals no new low-lying states.  We then turn out attention toward bases only containing five-quark operators, in order to allow any spectral strength that may have been otherwise overwhelmed by three-quark operators to participate in the analysis.  This approach has proved beneficial in the negative-parity nucleon channel~\cite{Kiratidis:2015vpa}.

The spectra obtained from a $6 \times 6$ analysis using the five-quark operators $\chi_{\pi{N}}$,
$\chi^{\prime}_{\pi{N}}$ and $\chi_{a_{0}N}$ are illustrated as basis
number four.  In this analysis we extract an energy
level that lies between the two previously observed states, but importantly no new low-lying state is extracted. In the final three columns we form $8 \times 8$ bases with the three possible
combinations of pairs of our five-quark operators.  Energy levels consistent
with those already extracted are observed, but crucially no new low-lying states are
found.  We note here that the spectrum extracted has some common features with the Hamiltonian effective field theory model in Ref.~\cite{Liu:2016uzk}.
We are now in a position to conclude that the introduction of local five-quark operators with a positive-parity meson contribution does not provide significant overlap with the low-lying finite-volume scattering states in this channel.  We also note that in bases four through seven, in which only five-quark operators were used, the ground state nucleon was extracted with a high degree of precision highlighting the meson-baryon cloud of the nucleon.

\section{Conclusions}
\label{sect:Conclusions}

In this exploratory investigation we introduced local five-quark interpolating fields with positive-parity meson contributions in the channel of the Roper resonance.  Motivated by success in the negative parity channel, the aim was to extract new low-lying states that had been
missed in previous calculations.

Following the construction of $a_{0}{N}$- and $\sigma{N}$-type operators, ratios of correlation functions were constructed to determine which operators held the greatest promise of revealing new states.  Informed by these ratios, correlation matrices were constructed from several different operator bases, and their associated spectra were produced.  

When compared to the spectra produced with solely three-quark operators, no new energy levels below the first excitation were extracted.  The local five-quark operators investigated were also found to possess a significant overlap with the ground state nucleon, as bases containing solely these operators enabled the extraction of the ground state with a high degree of precision.

We conclude that the low-lying finite-volume meson-baryon scattering states are not well localised.  Rather, the two-particle scattering states dominate so that the volume suppression associated with the overlap of scattering states with local operators prevents their extraction.  Our results strengthen the interpretation of the Roper as a coupled-channel dynamically-generated meson-baryon resonance that is not closely associated with conventional three-quark states.

\section*{Acknowledgments}
The authors thank the PACS-CS Collaboration for making the $2+1$ flavor gauge configurations available, as well as the
ongoing support of the ILDG.  This research was undertaken with the assistance of the University of Adelaide's Phoenix cluster and resources at
the NCI National Facility in Canberra, Australia.  These resources were provided through the
National Computational Merit Allocation Scheme, supported by the Australian Government and the
University of Adelaide Partner Share.  This research is supported by the Australian Research
Council through the ARC Centre of Excellence for Particle Physics at the Terascale (CE110001104),
and through Grants No.\ LE160100051, DP151103101 (A.W.T.), DP150103164, DP120104627 and LE120100181
(D.B.L.).

\bibliographystyle{JHEP} 
\bibliography{reference}

\end{document}